\documentclass[aps,prb,showpacs,reprint,twocolumn,amsmath,amssymb,nofootinbib,showkeys]{revtex4-2}
\usepackage{graphicx}
\usepackage{amssymb}
\usepackage{amsmath}
\begin{document}
\title{Optical and Hidden Transport Properties of BaFe$_{1.91}$Ni$_{0.09}$As$_{2}$ Film}

\author{Yu.A. Aleshchenko$^1$}
\email{*E-mail:aleshchenkoya@lebedev.ru}
\author{A.V. Muratov$^1$}
\author{G.A. Ummarino$^{2,3}$}
\author{S. Richter$^{4,5}$}
\author{A. Anna Thomas$^{4,5}$}
\author{R.H\"uhne$^4$}
\affiliation{$^1$ V.L. Ginzburg Center for High-Temperature Superconductivity and Quantum Materials, P.N. Lebedev Physical Institute, Russian Academy of Sciences, Leninskiy Prospekt 53, Moscow 119991, Russia}
\affiliation{$^2$ Istituto di Ingegneria e Fisica dei Materiali, Dipartimento di Scienza Applicata e Tecnologia, Politecnico di
Torino, Corso Duca degli Abruzzi 24, 10129 Torino, Italy}
\affiliation{$^3$ National Research Nuclear University MEPhI (Moscow Engineering Physics Institute), Kashirskoe hwy 31, Moscow 15409, Russia}
\affiliation{$^4$ Institute for Metallic Materials, Leibniz IFW Dresden, Helmholtzstrasse 20, Dresden 01069, Germany}
\affiliation{$^5$ School of Sciences, TU Dresden, 01062 Dresden, Germany}



\begin{abstract}
Optical spectroscopy was used to study the electrodynamics and hidden transport properties of a BaFe$_{1.91}$Ni$_{0.09}$As$_{2}$ thin superconducting film. We analyzed the normal state data using a Drude-Lorentz model with two Drude components: one narrow ($D_1$) and another broad one ($D_2$). In the superconducting state, two gaps with $2\Delta _{0}^{(2)}/k_BT_c=1.9$--2.0 and $2\Delta _{0}^{(1)}/k_BT_c=4.0$--4.3 are formed from the narrow component $D_1$ while the broad component $D_2$ remains ungapped. The calculated total DC resistivity of the film and the low-temperature scattering rate for the narrow Drude component show a hidden Fermi-liquid behavior. The change of total electron-boson coupling ($\lambda_{tot}$) and representative energy ($\Omega_{0}$) in the normal state with respect to the superconducting state is typical of other iron-based materials as well as high-temperature superconducting (HTSC) cuprates.
\end{abstract}

\maketitle

\section{INTRODUCTION}
Iron-based superconductors are among the most studied superconducting (SC) compounds due to their high critical temperatures $T_c$, large upper critical field $H_{c2}$, type-II nature and relatively low anisotropy~\cite{Jonhston 2010}. These characteristics make them promising candidates for various applications~\cite{Heindl 2014,Hosono 2018,Yao 2019}. Among these materials, doped BaFe$_2$As$_2$ (Ba122) compounds attracted special attention due to a comparatively simple technology of growth at ambient pressure, good quality of available single crystals and the fact that superconductivity can be induced in many ways (i.e., by application of external pressure, by substitution of each atomic site or by combination of both routes), resulting in phase diagrams that are comparable~\cite{Canfield 2010}. Previous studies of iron-arsenic materials recognized their multiband nature with hole and electron pockets at the center and corners of the Brillouin zone~\cite{Singh 2008,Fink 2009}. In this case, an analysis of optical data needs to include two different types of free-carrier contributions~\cite{Wu 2010}.

Many doped SC compounds of the Ba122-family can be grown as epitaxial thin films with decent quality~\cite{Hanish 2019}. However Fourier-transform infrared (FT-IR) spectroscopic studies of Ni-doped Ba122 epitaxial films have been reported only in~\cite{Yoon 2017}. The spectral range in this study was restricted to the interval from 50 to 7000~cm$^{-1}$ and nothing was said on the values of superconducting gaps. Moreover, even for a bulk Ba(Fe,Ni)$_2$As$_2$ data on the SC gap is scarce~\cite {Chi 2009,Ding 2009,Gong 2010,Dressel 2011,Kuzmicheva 2016,Kuzmicheva 2018}.

In our recent paper~\cite{Ummarino 2020}, we studied the optical properties of the BaFe$_{1.91}$Ni$_{0.09}$As$_2$ film by THz spectroscopy within the range of 10--50~cm$^{-1}$. The experimental data was analyzed within a simple three-band Eliashberg model, where the mechanism of superconducting coupling is mediated by antiferromagnetic spin fluctuations, whose characteristic energy  $\Omega _0$ scales with $T_c$ according to the empirical law $\Omega _{0} = 4.65k_{B}T_c$, and with a total electron-boson coupling strength $\lambda_{tot} = 2.17$.

In the present paper, we studied the optical and hidden transport properties of the similar BaFe$_{1.91}$Ni$_{0.09}$As$_2$ film by the combination of FT-IR spectroscopy and ellipsometry. Optical spectroscopy provides rich information on the bulk electronic properties of superconductors. It probes the energy, scattering and symmetry of the SC gaps and sum rules can be used to determine the SC condensate density. This data provides a deep insight into the pairing mechanisms in iron-based superconductors.

\section{EXPERIMENT}
The nearly optimum electron doped BaFe$_{1.91}$Ni$_{0.09}$As$_{2}$ film with an approximate thickness of $120\pm 10$~nm and an area of $10\times 10$~mm$^2$ was grown by pulsed laser deposition (PLD) on polished (001) CaF$_2$ substrate. More details on the sample preparation can be found in~\cite{Richter 2017,S Richter 2017,Shipulin 2018}. Electronic transport measurements of the film were performed using a standard four-probe configuration in a physical property measurement system (PPMS, Quantum Design). Figure~1 shows the dc in-plane resistivity of the BaFe$_{1.91}$Ni$_{0.09}$As$_{2}$ film as a function of temperature (black open circles). The resistivity $\rho (T)$ is characterized  by a sharp SC transition at $T_c\simeq 20$~K and a tendency to saturation at room temperature. The IR reflectivity spectra were measured near normal incidence in the spectral range of 25--10000~cm$^{-1}$ (400--1~$\mu $m) at various temperatures between 5 K and 300 K using a conventional FT-IR spectrometer (IFS 125HR, Bruker) equipped with a Si-bolometer for measurements in the far-IR range and Konti Spectro A continuous-flow cryostat. A freshly evaporated gold mirror was used as a reference with the necessary correction for the spectrum of gold. In order to determine the properties of the film, we measured the optical parameters of a bare CaF$_2$ substrate beforehand over the entire frequency and temperature range (see Supplemental Material for more details on the experimental procedure and data analysis). Additionally, variable-angle spectroscopic ellipsometry was perfomed from 4000 to 50000 cm$^{-1}$ (2.5~$\mu $m---200~nm) with a Woollam VASE ellipsometer in a high-vacuum Janis CRF 725V cryostat (4.2--300~K). The ellipsometry data was converted to normal-incidence reflectivity data. In this connection, it should be mentioned that the near-normal incidence reflectivity probes only the $ab$-plane dielectric function $\varepsilon _{ab}$, while ellipsometry data is  taken at larger angles of incidence (usually at 65$^\circ $--75$^\circ $), i.e. the $c$-axis dielectric function $\varepsilon _c$ of the film cannot be ignored, in general. However, in contrast to  superconducting cuprates, the $ab$-plane and $c$-axis optical conductivities in the 122 iron arsenides differ by only 20-30\%~\cite{Cheng 2011,Chen 2010}. In this case, it was shown in~\cite{Xing 2016} for Ca$_{0.8}$La$_{0.2}$Fe$_2$As$_2$ that the pseudodielectric function taken directly from ellipsometry data is an accurate representation of the $ab$-plane dielectric function. In fact, the larger $\varepsilon _{ab}$ the smaller the influence of the $c$-axis optical constants on the pseudodielectric function. This correlation allows to calculate the $ab$-plane reflectivity of our film from ellipsometry data.
\begin{figure}
\includegraphics{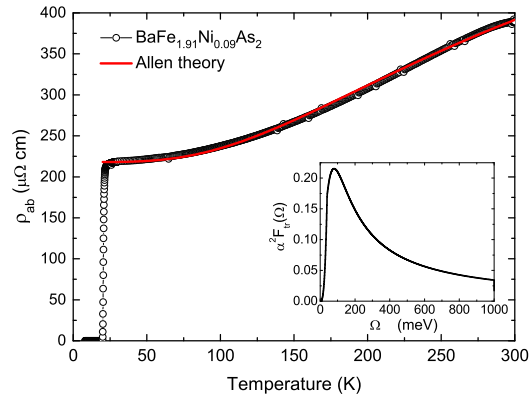}
\caption{(Color online) Temperature dependence of the resistivity for the BaFe$_{1.91}$Ni$_{0.09}$As$_{2}$ film (black open circles). The solid red curve represents the fit obtained with the Allen theory (see Section~\ref{Resistivity} for more details). The inset shows the spectral function of the antiferromagnetic spin fluctuations in the normal state (see Section~\ref{Resistivity}).}
\end{figure}

\section{EXPERIMENTAL RESULTS}
Figure 2(a) shows the broad band reflectivity spectra $R(\nu)$ of the BaFe$_{1.91}$Ni$_{0.09}$As$_{2}$ film at various temperatures together with the fit based on the Drude-Lorentz analysis (see below). The reflectivity exhibits a metallic response for both frequency and temperature. The reflectivity curves almost overlap with each other above 100~cm$^{-1}$ at temperatures of 5--100 K. Above 5700 cm$^{-1}$ the difference between the spectra, including those taken at 100--300 K, is hardly discernible. However, a sudden upturn $R(\nu)$ develops below $T_c$ at frequencies as low as 60 cm$^{-1}$ (7.4 meV). This is a strong indication for the formation of an SC energy gap due to the pairing of electrons. The decrease of $R(\nu)$ becomes steep above 300 cm$^{-1}$. In addition, our data contains two sharp features at $\sim 94$ and $\sim 260$~cm$^{-1}$ which correspond to the contributions of phonons of the CaF$_2$ substrate and symmetry-allowed $ab$-plane IR-active $E_u$ phonons of the film~\cite{Akrap 2009}.
\begin{figure}
\includegraphics{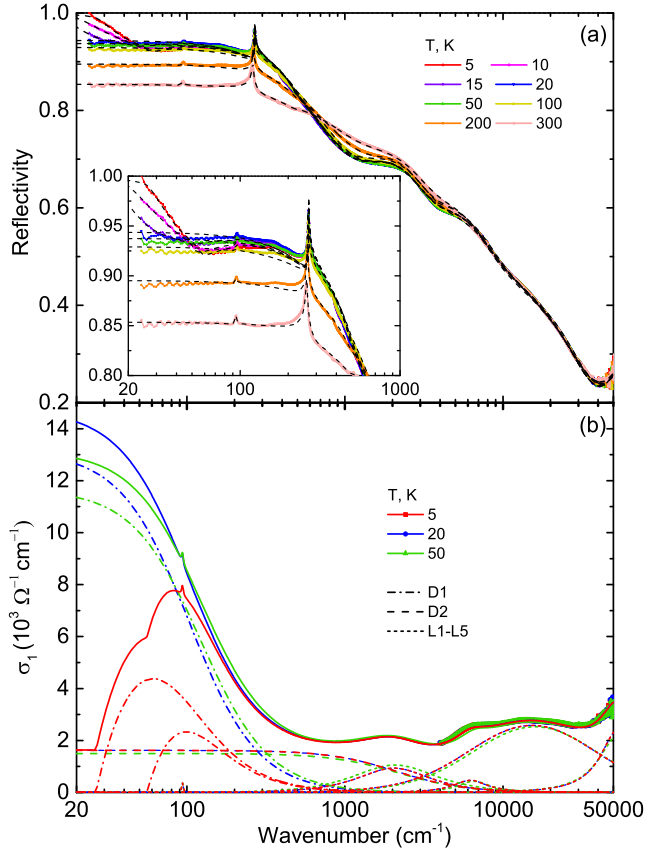}
\caption{(Color online) (a) The measured reflectivity spectra of the BaFe$_{1.91}$Ni$_{0.09}$As$_{2}$ film at various temperatures together with the fit (dashed lines). In the inset we display magnified views of reflectivity spectra for the low frequency region. (b) Three representative Drude-Lorentz fits and optical conductivity data for the BaFe$_{1.91}$Ni$_{0.09}$As$_{2}$ film at temperatures of 5, 20 and 50 K, respectively.}
\end{figure}

We applied a full two-layer Drude-Lorentz modeling of the film/substrate system with a finite substrate thickness to the analysis of the experimental reflectivity together with the dielectric permittivity and optical conductivity (from the ellipsometry data) of the BaFe$_{1.91}$Ni$_{0.09}$As$_{2}$ film. We used a two-Drude approach proposed earlier in~\cite{Wu 2010} to study properties of multiband iron pnictides. In this case, the optical conductivity in the normal state is modeled by two Drude components, one narrow ($D_1$) and another broad one ($D_2$), and by a set of Lorentz components representing the interband transitions. An alternative approach with the replacement of the broad Drude term by an overdamped Lorentzian is discussed in the Supplemental Material. The complex dielectric function $\tilde\varepsilon (\omega)=\varepsilon _1(\omega)+i\varepsilon _2(\omega)$ can be written as
$$
\tilde\varepsilon (\omega)=\varepsilon _{\infty }-\sum _{i=1,2}\frac{\omega _{Di,p}^2}{\omega (\omega +i\gamma _{Di})}+\sum _j\frac{\omega _{j,p}^2}{\omega _j^2-\omega ^2-i\gamma _j\omega},
$$
where $\varepsilon _\infty $ is the background dielectric function, which comes from contribution of the high frequency absorption, $D_1$ ($D_2$) stands for the narrow (broad) Drude component, $\omega _{Di,p}$ is the Drude plasma frequency, $\gamma _{Di}$ is the (average) elastic scattering rate among free charge carriers, $\omega _{j,p}$, $\omega _j$, and $\gamma _j$ are the plasma frequency, the center frequency, and the width of the $j$th Lorentz component, respectively. The optical conductivity can be related to the dielectric function as $\tilde\sigma (\omega )=\sigma _1(\omega)+i\sigma _2(\omega )=i\omega [\varepsilon _\infty -\tilde\varepsilon (\omega )]/4\pi $.

Below $T_c$, the Drude term in the Drude-Lorentz model should be replaced by the Zimmermann term~\cite{Zimmermann 1991}, which generalizes the standard BCS Mattis-Bardeen model~\cite{Tinkham 1996,Dressel 2002} to arbitrary $T$ and $\gamma $ values, with the inclusion of two additional parameters, the SC gap $\Delta $ and the ratio $T/T_c$.

In Fig.~2(b) we show the results of the modeling for the real part of the optical conductivity spectrum at various temperatures based on the Drude-Lorentz analysis. The noisy curves in the range of 4000--50000~cm$^{-1}$ are the optical conductivities calculated directly from the ellipsometry data. The CaF$_2$ substrate has the strongest phonon peak at $\sim 260$~cm$^{-1}$. It was not included into the film model, which, nevertheless, provided a good quality of the fit. For the same reason, the phonon peak at $\sim 94$~cm$^{-1}$ was included only into the film model. In the normal state, the optical conductivity is decomposed into a narrow $D_1$ and a broad $D_2$ Drude terms as well as a set of Lorentzians representing interband transitions. The observation of narrow and broad terms in the two-Drude analysis is in agreement with other optical studies, and appears to be a general feature for most of the iron-based materials having multiband structure~\cite{Wu 2010}. As temperature decreases, the $D_1$ component increases and gets narrower while the $D_2$ component as well as Lorentz contributions remain virtually unchanged.

Upon passing the superconducting transition a gaplike structure is formed in the narrow Drude term. Assuming two isotropic SC gaps both opening simultaneously below $T_c$, we obtain a better description of the low-frequency optical conductivity $\sigma _1$ with $2\Delta _{0}^{(1)}=56$--60~cm$^{-1}$ (6.9--7.4~meV) and $2\Delta _{0}^{(2)}=26$--27~cm$^{-1}$ (3.2--3.4~meV). Forty five percent of the initial conductivity is due to the larger gap and fifty five percent of the conductivity is accounted for by the smaller one. The gap values obtained from our study are $2\Delta _{0}^{(1)}/k_BT_c=4.0$--4.3 and $2\Delta _{0}^{(2)}/k_BT_c=1.9$--2.0, respectively. These values are consistent with the SC gaps of BaFe$_{1.9}$Ni$_{0.1}$As$_2$ single crystals determined from specific heat measurements as well as directly by intrinsic multiple Andreev reflections effect (IMARE) spectroscopy~\cite{Kuzmicheva 2016,Kuzmicheva 2018}.
A three-band Eliashberg model used for the analysis of our previous THz spectroscopy data for the similar BaFe$_{1.91}$Ni$_{0.09}$As$_2$ film ($T_c=19.6$~K)~\cite{Ummarino 2020} produces quite different values of the SC gaps, $2\Delta _1/k_BT_c=5.1$, $2\Delta _2/k_BT_c=3.3$, and $2\Delta _2/k_BT_c=7.0$. Such a difference can be related to the slightly different $T_c$ of our films as well as the coupling constants for Ba(Fe,Co)$_2$As$_2$ used in previous calculations.

It is seen from Fig.~2(b) that the broad Drude term is not gapped at least  in the range of our measurements. For the case that the coherent transport in the normal state arises mainly from the electron pocket, the observed two energy scales for the superconducting energy gaps might be due to a  single anisotropic $s$-wave gap at the electron pocket (see the results for Co-Ba122 having very similar electronic structure~\cite{Tu 2010}). This picture is consistent with a $s_\pm $ symmetry of the order parameter.

It should be noted that the use of other models, e.g., with one gap either in the $D_1$ or $D_2$ term or two gaps belonging to the different terms or to the $D_2$ term does not provide an adequate description of the experimental data. However, the replacement of the broad Drude term by an overdamped Lorentzian also provides a good fit to the reflectivity spectra of the BaFe$_{1.91}$Ni$_{0.09}$As$_2$ film in the normal state [see Fig.~S4(a) in the Supplemental Material]. In the SC state, a simple single-band theory fails to describe the low-frequency reflectivity spectrum [dash-dot curve in the inset of Fig.~S4(a)]. The consideration of two SC gaps both opening in a single band or in the different bands having similar parameters provides the best fit with the gap magnitudes $2\Delta _0^{(1)}=51.5$~cm$^{-1}$ and $2\Delta _0^{(2)}=23.6$~cm$^{-1}$ [Fig.~S4(b) in the Supplemental Material]. These values are very close to those obtained above with a two-Drude model, which confirms the reliability of these results.

To see the role of impurities in the mechanism of superconductivity, it is instructive to compare the magnitudes of SC gaps obtained in this study for BaFe$_{1.91}$Ni$_{0.09}$As$_2$ with those extracted from IR for other electron doped Ba122 systems such as Co-Ba122 and Pt-Ba122 with the similar levels of doping. Since Co donates only one additional 3d electron compared to Fe, while Ni has two 3d electrons more, one should use the composition close to BaFe$_{1.82}$Co$_{0.18}$As$_2$ for comparison (Table I). In the case of BaFe$_{1.9}$Pt$_{0.1}$As$_2$, optical conductivity data is consistent with three nodeless energy gaps in the SC state~\cite{Xing 2018}. From Table~I we may conclude that the same quantity of similar dopant produces analogous effect on the $T_c$ and SC gap magnitudes in electron-doped Ba122 ferropnictides. This means that the mechanism of superconductivity in these materials is robust against the change of dopants with similar properties and not prone to fine tuning.
\begin{table*}
\begin{center}
\caption{Some SC papameters for electron-doped Ba122 ferropnictides with the similar levels of doping.}
\begin{tabular}{|c|c|c|c|c|c|c|c|}
\hline
Compound & $T_c$, K & $\Delta _<$, & $\Delta _>$, & $2\Delta _</k_BT_c$ & $2\Delta _>/k_BT_c$ & Method & Reference\\
& & meV & meV & & & & \\
\hline
BaFe$_{1.91}$Ni$_{0.09}$As$_2$ & 20 & 1.6--1.7 & 3.45--3.7 & 1.9--2.0 & 4.0--4.3 & Optical & ---\\
& & & & & & conductivity, & \\
& & & & & & this study & \\
\hline
BaFe$_{1.8}$Co$_{0.2}$As$_2$ & 20 & 1.85 & 3.95 & 2.15 & 4.6 & Optical & \cite{Maksimov}\\
& & & & & & conductivity & \\
\hline
BaFe$_{1.9}$Pt$_{0.1}$As$_2$ & 23 & 1.95 & 3.6, & 1.97 & 3.63, & Optical & \cite{Xing 2018}\\
& & & 5.4 & & 5.44 & conductivity & \\
\hline
\end{tabular}
\end{center}
\end{table*}

\begin{figure}
\includegraphics{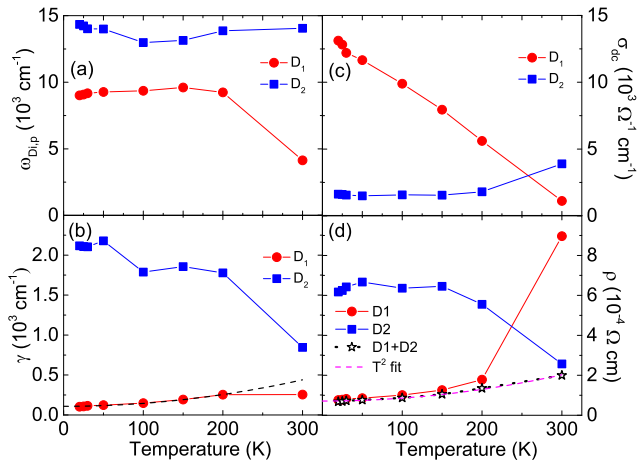}
\caption{(Color online) The fitting parameters $\omega _{Di,p}$ (a) and $\gamma _{Di}$ (b) of two Drude modes $D_i$ ($i=1, 2)$ of the BaFe$_{1.91}$Ni$_{0.09}$As$_{2}$ film. We also show the calculated DC conductivities $\sigma _{dc,i}(T)$ (c) and DC resistivity $\rho (T)$ (d) including the total resistivity as functions of temperature. The dashed lines are $T^2$ fits to the scattering rate for the narrow Drude component and to the total DC resistivity.}
\end{figure}

In Fig. 3 are shown the temperature dependences of the parameters of two Drude modes for the BaFe$_{1.91}$Ni$_{0.09}$As$_{2}$ film in the normal state. The parameters are the plasma frequency ($\omega _{Di,p}$), the static scattering rate ($\gamma $), the DC conductivity ($\sigma _{DC}$), and the DC resistivity $\rho $ of the two Drude modes. One can see that $\omega _{D1,p}$, $\gamma _{D2}$, and $\rho _{D2}$ show some temperature dependences as opposed to their invariability found for Ba$_{0.6}$K$_{0.4}$Fe$_2$As$_2$ single crystals~\cite{Dai 2013}, Ni-Ba122 single crystals~\cite{Lee 2015}, and BaFe$_{1.9}$Ni$_{0.1}$As$_2$ film~\cite{Yoon 2017}. However, at least below 200~K these parameters are nearly temperature independent. Moreover, the Drude plasma frequency $\omega _{D,p}=\sqrt{4\pi ne^2/m^*}$ ($n$ is the carrier concentration, $m^*$ is an effective mass) deduced from the one Drude--one Lorentzian model is also temperature independent as is evident from Fig.~S3(a) of the Supplemental Material. This indicates that the band structure and $n/m^*$ of BaFe$_{1.91}$Ni$_{0.09}$As$_2$ do not change noticeable with temperature. On the contrary, the scattering rate data for the narrow Drude contribution indeed demonstrates a quadratic dependence at low temperatures as is evident from Fig.~3b with the quadratic function shown by the dashed line. The total DC resistivity estimated from the optical data can be calculated using the expression $1/\rho _{D1+D2}=1/\rho _{D1}+1/\rho _{D2}$. It shows a $T^2$ behavior as can be seen from Fig.~3(d). Such a behavior is also characteristic for $\gamma _{D0}(T)$ and $\rho (T)$ obtained within a one Drude--one Lorentzian model described in Supplemental Material [see Fig.~S3(b),(d)]. This is a further proof for the validity of the obtained results. A quadratic temperature dependence is expected for electron-electron scattering which is supposed to be dominant for correlated electron systems at low temperatures and subject to Landau's theory of a Fermi liquid~\cite{Abrikosov 1963,Pines 1966}. Such a behavior was previously observed for BaFe$_{1.9}$Ni$_{0.1}$As$_2$ PLD film~\cite{Yoon 2017} as well as for  BaFe$_{1.9}$Ni$_{0.1}$As$_2$ and BaFe$_{1.84}$Co$_{0.16}$As$_2$ single crystals~\cite{Lee 2015,Barisic 2010}, while for hole-doped Ba$_{0.6}$K$_{0.4}$Fe$_2$As$_2$ a $T$-linear behavior was found~\cite{Dai 2013}. This indicates that the electron- and hole-doped samples show different hidden $D_1$ transport properties.

The total DC resistivity estimated from the optical data shows a similar temperature dependence but some difference in magnitude as compared with the transport resistivity data. We think that the difference might originate from the uncertainties of the resistivity data obtained by two different measurement methods. In the following, we want to discuss the resistivity in more detail.

\section{MODEL FOR THE RESISTIVITY IN A MULTIBAND METAL}
\label{Resistivity}
The resistivity in a multiband case can then be obtained by extending the single band case~\cite{Allen, Grimvall} and considering the contribution of all the different channels:
\begin{equation}
	\frac{1}{\rho(T)}=\frac{\varepsilon_0(\hbar\omega_{p})^{2}}{\hbar}\sum_{i=1}^N
						\frac{(\omega_{p,i}/\omega_{p})^2}{\gamma_i+W_i(T)},
\label{rho}
\end{equation}
where $N$ is the total number of the different carriers considered, $\omega_{p,i}$ is the bare plasma frequency of the $i$-band, $\omega_{p}$ is the total bare plasma frequency  and
\begin{equation}
	W_i(T)=4\pi k_BT\int_0^\infty d\Omega
			\left[\frac{\hbar\Omega/2k_BT}{\sinh\big(\hbar\Omega/2k_BT\big)}\right]^2
			\frac{\alpha_{tr,i}^2F_{tr}(\Omega)}{\Omega},\\
\label{W1}
\end{equation}
with $\gamma_{i}=\sum_{j=1}^N\Gamma_{ij}+\Gamma^{M}_{ij}$ is the sum of the inter- and intra-band non magnetic and magnetic impurity scattering rates present in the Eliashberg equations and
\begin{equation}
\alpha_{tr,i}^{2}F_{tr}(\Omega)=\sum_{j=1}^N\alpha_{tr,ij}^{2}F_{tr}(\Omega),
\end{equation}
where $\alpha^2_{tr,ij}(\Omega)F_{tr}(\Omega)$ are the transport spectral functions related to the Eliashberg functions~\cite{Allen}.\\
We can define a normalized spectral functions $\alpha_{tr,ij}^{2}F^{n}_{tr}(\Omega)$ that has an electron-boson coupling constant $\lambda=2\int_{0}^{+\infty}\frac{\alpha^{2}F(\Omega)}{\Omega}d\Omega$ equal to $1$ which results in
$\alpha_{tr,ij}^{2}F_{tr}(\Omega)=\lambda_{tr,ij}\alpha_{tr,ij}^{2}F^{n}_{tr}(\Omega)$. \\

To write the reciprocal resistivity as a sum of N different contributions is an approximation that was discussed already in the past in ref.~\cite{Allen 1981}. This approximation works, if the material is not in a strong disorder regime. This means in particular for the temperature range of interest the resistivity has to be less than a maximum value $\rho _{max}=4\pi\hbar E_F/[\varepsilon _0(\hbar\omega_p)^2]$, where $\varepsilon _0$ is the vacuum dielectric constant, $E_F=0.6$~eV~\cite{Golubov 2011} is the Fermi energy, and $\hbar\omega _p=2101$~meV [since from our optical measurements $\hbar\omega _{p,1}=1118$~meV and $\hbar\omega _{p,2}=1779$~meV (see the values of $\omega _{p,1}$ and $\omega _{p,2}$ at 20~K in Fig.~3a), and $\omega _p^2=\omega _{p,1}^2+\omega _{p,2}^2$]. If we take these, we find $\rho _{max}=13$~m$\Omega\cdot $cm, which is significantly larger than the measured resistivity for our film; therefore we can use this approximation. As in the superconducting state, we set all spectral functions to be equal and they differ just for a scaling factor, the coupling constant (this is a very good approximation, especially if the coupling is mediated mainly by spin fluctuations), then
$\alpha_{tr,i}^{2}F_{tr}(\Omega)=\lambda_{tr,i}\alpha_{tr}^{2}F^{n}_{tr}(\Omega)$ and $\lambda_{tr,i}=\sum_{j=1}^N \lambda_{tr,ij}$. The specific shape of the spectral function for the antiferromagnetic spin fluctuation in the normal state is standard ~\cite{resumma}. The shape and position of the peak are the same for the spectral functions in all bands. They are just rescaled for having different $\lambda _{ij}$ (electron-boson coupling constant).

The Fermi surface consists of several sheets, therefore a multi-band model is in principle required to explain superconducting and normal state properties. In order to study the electrical resistivity, we grouped the hole and the electron bands, i.e. a model containing only two different kind of carriers is used.
Considering the fact that the electron-phonon coupling in all iron-based superconductors is weak~\cite{Lilia}, it is logic to consider another mechanism, which contributes to the transport properties. Knowing the basic superconducting properties of iron pnictides, antiferromagnetic spin fluctuations are the best candidates to play the role of the principal actor also in the normal state. The transport spectral functions are similar to the standard Eliashberg functions~\cite{Allen}, the only difference between superconducting and normal state is the behavior for $\Omega\rightarrow 0$, where the transport function behaves like $\Omega^3$ instead of $\Omega$. Therefore, the condition $\alpha^2_{tr}(\Omega)F_{tr}(\Omega)\propto\Omega^3$ should be imposed in the range $0 <\Omega< \Omega_{D}$, with $\Omega_{D}\approx\Omega_{0}/10$ where $\Omega_{0}$ is the representative bosonic energy~\cite{Allen}. Then
$\alpha^2_{tr}(\Omega)F'_{tr}(\Omega)=
b_{i}\Omega^{3}\vartheta(\Omega_{D}-\Omega)
+c_{i}\alpha^2_{tr}(\Omega)F''_{tr}(\Omega)\vartheta(\Omega-\Omega_{D})$
and the constants $b_{i}$ and $c_{i}$ have been fixed by requiring the continuity in $\Omega_{D}$ and the normalization. This function is shown in the inset of Fig.~1.

In order to describe the coupling with spin fluctuations, we choose for the $\alpha^2_{tr}(\Omega)F''_{tr}(\Omega)$ the functional form of the theoretical antiferromagnetic spin fluctuations function in the normal state~\cite{Popovich} that reproduces the experimental data concerning the normal state dynamical spin susceptibility~\cite{Inosov}

\begin{equation}
\alpha_{tr}^{2}F''(\Omega)\propto\frac{\Omega_{0}\Omega}{\Omega^{2}+\Omega_{0}^{2}}\vartheta(\Omega_{c}-\Omega),
\label{a2F}
\end{equation}
where $\Omega_{c}$ is a cut-off energy (in these calculations \mbox{$\Omega_{c}=1$ eV}) and $\Omega_{0}$ is the energy of the peak.

As mentioned above, the electronic structure in these compounds consists of hole and electron pockets. In order to keep the number of free parameters as low as possible and taking into account that just one carrier is not enough, we consider a model containing only two different kinds of carriers. Within this model the electron-boson coupling constants $\lambda_{1,tr}$ and $\lambda_{2,tr}$, the impurities contents $\gamma_1$ and $\gamma_2$, the plasma energies $\omega_{p,1}$ and $\omega_{p,2}$, the representative energy $\Omega_{0}$ of the transport electron-boson spectral functions and the energy $\Omega_{D}$ are free parameters. In the Co doped case, ARPES and de Haas-van Alphen data suggest that the transport is drawn mainly by the electronic bands and that the hole bands are characterized by a weaker mobility~\cite{Maksimov}. This means that within our model the impurities are mostly concentrated in the hole band (indicated by the index 2), i.e. $\gamma_2 \gg \gamma_1$, while the transport coupling is much higher in the electron band (indicated by the index 1); then, at least as a first approximation, $\lambda_{2,tr}$ can be fixed to be zero. In this way one contribution results to be temperature independent and the other changes the slope of the resistivity with the temperature. The values of $\gamma_1$ and $\gamma_2$ are connected with the experimental value of $\rho(T=0)=218$ $\mu\Omega $$\,$cm so just one is free. We take the total plasma energy $\omega_{p}=2101$~meV from our optical measurements as $\omega_{p,1}=1118$~meV ($\omega_{p,2}=1779$~meV) and $\gamma_2=262$ meV ($\gamma_1=50$ meV), all at $T=20$ K, so at the end we have 3 free parameters. We find $\lambda_{1,tr}=1.59$, $\Omega_{0}=80$ meV and we choose, as in the Co doped case~\cite{resumma}, $\Omega_{D}=\Omega_{0}/2 > \Omega_{0}/10$. The value of parameters found are similar to the Co doped case and also the total coupling $\lambda_{tr}=0.51$ is small~\cite{resumma}. The final result is good as it is shown in Fig.~1 (red line).

We note that there is a big increase for the typical energy $\Omega_0$ of the electron-boson coupling from the superconducting to the normal state in agreement with inelastic neutron scattering experimental data~\cite{Inosov}. Moreover, the electron-boson spectral density extracted using Eliashberg equations from optical conductivity of Co doped BaFe$_2$As$_2$~\cite{Carbi} shows exactly the same behavior and is another evidence that in iron-based superconductors spin fluctuations strongly couple to the charge carriers and mediate superconductivity. The fit of the dc resistivity shows also a significant decrease of the coupling constant $\lambda_{tr}$ in relation to the superconducting state ($\lambda_{tot}$). It was shown for simple metals that $\lambda_{tr}\le\lambda_{tot}$ while in strongly correlated materials like HTSC cuprates holds $\lambda_{tr}\ll\lambda_{tot}$~\cite{Maxi}. This behavior is also typical of other iron-based materials~\cite{resumma,Golubov 2011} and may have its origin in strong electronic correlations~\cite{Maxi}. This indicates that for {\it ab initio} calculations of the physical properties of these materials electronic correlations cannot be neglected.

\section{CONCLUSION}
In conclusion, we studied the optical and hidden transport properties of the nearly optimum doped BaFe$_{1.91}$Ni$_{0.09}$As$_2$ SC film. A two-Drude model was found to describe successfully the optical properties of the film studied. In the superconducting state two gaps with $2\Delta _{0}^{(2)}/k_BT_c=1.9$--2.0 and $2\Delta _{0}^{(1)}/k_BT_c=4.0$--4.3 are formed from the narrow Drude component $D_1$ while the broad component $D_2$ remains ungapped. An alternative one Drude--one Lorentzian model gives almost similar values for the SC gaps but admits their opening in different bands having similar parameters. The temperature dependences of the model parameters in the normal state are examined within both models. A hidden Fermi-liquid behavior found earlier for optimally doped Ba(Fe,Ni)$_2$As$_2$ PLD film~\cite{Yoon 2017} and single crystal~\cite{Lee 2015} is confirmed. From the resistivity measurement as a function of temperature we realize that the electron-boson coupling constant is strongly reduced and the representative boson energy is strongly increased as it happens with the other iron-based materials~\cite{resumma,Golubov 2011} and HTSC cuprates~\cite{Maxi}. The similar behavior probably results from the fact that the mechanism responsible for superconductivity is, perhaps, similar for HTSC cuprates and iron-based materials.

\section{ACKNOWLEDGMENTS}
The work of Yu.A.A. and A.V.M. is carried out within the state assignment of the Ministry of Science and Higher Education of the Russian Federation (theme "Physics of high-temperature superconductors and novel quantum materials", No. 0023-2019-0005). G.A.U. acknowledges support from the MEPhI Academic Excellence Project (Contract No. 02.a03.21.0005). R.H., S.R. and A.A.T. acknowledge the financial support by the German Research Foundation (DFG) within the framework of the research training group GRK1621. The spectroscopic measurements were done using research equipment of the Shared Facilities Center at LPI.\\
\\


\begin{thebibliography}
\small{
%
\bibitem{Jonhston 2010} D. C. Johnston, \textit{Advances in Physics} \textbf{59}, 803-1061 (2010).
%
\bibitem{Heindl 2014} S. Haindl, M. Kidszun, S. Oswald, C. Hess, B. B\"uchner, S. K\"olling, L. Wilde, T. Thersleff, V. V. Yurchenko, M. Jourdan, H. Hiramatsu, and H. Hosono, \textit{Rep. Prog. Phys.} \textbf{77} 046502 (2014).
%
\bibitem{Hosono 2018} Hideo Hosono, Akiyasu Yamamoto, Hidenori Hiramatsu, and Yanwei Ma, \textit{Materials Today} \textbf{21} 278-302 (2018).
%
\bibitem{Yao 2019} Chao Yao, Yanwei Ma, \textit{Supercond. Sci. Technol.} \textbf{32} 023002 (2019).
%
\bibitem{Canfield 2010} P. C. Canfield and S. L. Bud'ko, \textit{Annu. Rev. Condens. Matter Phys.} \textbf{1}, 27-50 (2010).
%
\bibitem{Singh 2008} D. J. Singh, \textit {Phys. Rev. B} \textbf {78}, 094511 (2008).
%
\bibitem{Fink 2009} J. Fink, S. Thirupathaiah, R. Ovsyannikov, H. A. D\"urr, R. Follath, Y. Huang, S. de Jong, M. S. Golden, Y.-Z. Zhang, H. O. Jeschke, R. Valent\'i, C. Felser, S. Dastjani Farahani, M. Rotter, and D. Johrendt, \textit {Phys. Rev. B} \textbf {79}, 155118  (2009).
%
\bibitem{Wu 2010} D. Wu, N. Bari\u{s}i\'{c}, P. Kallina, A. Faridian, B. Gorshunov, N. Drichko, L. J. Li, X. Lin, G. H. Cao, Z. A. Xu, N. L. Wang, and M. Dressel, \textit {Phys. Rev. B} \textbf {81}, 100512(R) (2010).
%
\bibitem{Hanish 2019} J. H\"anisch, K. Iida, R. H\"uhne, and C. Tarantini, \textit {Supercond. Sci. Technol.} \textbf {32}, 093001 (2019).
%
\bibitem{Yoon 2017} Sejun Yoon, Yu-Seong Seo, Seokbae Lee, Jeremy D. Weiss, Jianyi Jiang, MyeongJun Oh, Jongmin Lee, Sehun Seo, Youn Jung Jo, Eric E. Hellstrom, Jungseek Hwang, and Sanghan Lee, \textit {Supercond. Sci. Technol.} \textbf {30}, 035001 (2017).
%
\bibitem{Chi 2009} Songxue Chi, Astrid Schneidewind, Jun Zhao, Leland W. Harriger, Linjun Li, Yongkang Luo, Guanghan Cao, Zhu'an Xu, Micheal Loewenhaupt, Jiangping Hu, and Pengcheng Dai, \textit {Phys. Rev. Lett.} \textbf {102}, 107006 (2009).
%
\bibitem{Ding 2009} L. Ding, J. K. Dong, S. Y. Zhou, T. Y. Guan, X. Qiu, C Zhang, L. J. Li, X. Lin, G. H. Cao, Z. A. Xu, and S. Y. Li, \textit {New J. Phys.} \textbf {11}, 093018 (2009).
%
\bibitem{Gong 2010} Y. Gong, W. Lai, T. Nosach, L. J. Li, G. H. Cao, Z. A. Xu, and Y. H. Ren, \textit {New J. Phys.} \textbf {12}, 123003 (2010).
%
\bibitem{Dressel 2011} Martin Dressel, Dan Wu, Neven Bari\u{s}i\'{c}, and Boris Gorshunov, \textit {Journal of Physics and Chemistry of Solids} \textbf {72}, 514-518 (2011).
%
\bibitem{Kuzmicheva 2016} T. E. Kuzmicheva, V. A. Vlasenko, S. Yu. Gavrilkin, S. A. Kuzmichev, K. S. Pervakov, I. V. Roshchina, and V. M. Pudalov, \textit {J. Supercond. Nov. Magn.} \textbf {29}, 3059-64 (2016).
%
\bibitem{Kuzmicheva 2018} T. E. Kuzmicheva, S. A. Kuzmichev, A. V. Sadakov, S. Yu. Gavrilkin, A. Yu. Tsvetkov, X. Lu, H. Luo, A. N. Vasiliev, V. M. Pudalov, Xiao-Jia Chen, and Mahmoud Abdel-Hafiez, \textit {Phys. Rev. B} \textbf {97}, 235106 (2018).
%
\bibitem{Ummarino 2020} G. A. Ummarino, A. V. Muratov, L. S. Kadyrov, B. P. Gorshunov, S. Richter, A. Anna Thomas, R. H\"uhne, and Yu. A. Aleshchenko, \textit {Supercond. Sci. Technol.} \textbf {33}, 075005 (2020).
%
\bibitem{Richter 2017} Stefan Richter, Saicharan Aswartham, Aurimas Pukenas, Vadim Grinenko, Sabine Wurmehl, Werner Skrotzki, Bernd B\"uchner, Kornelius Nielsch, and Ruben H\"uhne, \textit {IEEE Transactions on Applied Superconductivity} \textbf {27}, 7300304 (2017).
%
\bibitem{S Richter 2017} Stefan Richter, Fritz Kurth, Kazumasa Iida, Kirill Pervakov, Aurimas Pukenas, Chiara Tarantini, Jan
    Jaroszynski, Jens H\"anisch, Vadim Grinenko, Werner Skrotzki, Kornelius Nielsch, and Ruben H\"uhne, \textit{Appl. Phys. Lett.} \textbf{110}, 022601 (2017).
%
\bibitem{Shipulin 2018} I. Shipulin, S. Richter, A. Anna Thomas, M. Brandt, S. Aswartham, and R. H\"uhne, \textit {Mater. Res. Express} \textbf {5}, 126001 (2018).
%
%
\bibitem{Cheng 2011} B. Cheng, Z. G. Chen, C. L. Zhang, R. H. Ruan, T. Dong, B. F. Hu, W. T. Guo, S. S. Miao, P. Zheng, J. L. Luo, G. Xu, P. Dai, and N. L. Wang, \textit{Phys. Rev. B} \textbf{83}, 144522 (2011).
%
\bibitem{Chen 2010} Z. G. Chen, T. Dong, R. H. Ruan, B. F. Hu, B. Cheng, W. Z. Hu, P. Zheng, Z. Fang, X. Dai, and N.L.Wang, \textit{Phys. Rev. Lett.} \textbf{105}, 097003 (2010).
%
\bibitem{Xing 2016} Z. Xing, T. J. Huffman, P. Xu, A. J. Hollingshad, D. J. Brooker, N. E. Penthorn, M. M. Qazilbash, S. R. Saha, T. Drye, C. Roncaioli, and J. Paglione, \textit{Phys. Rev. B} \textbf{94}, 064514 (2016).
%
\bibitem{Akrap 2009} A. Akrap, J. J. Tu, L. J. Li, G. H. Cao, Z. A. Xu, and C. C. Homes, \textit {Phys. Rev. B} \textbf {80}, 180502(R) (2009).
%
\bibitem{Zimmermann 1991} W. Zimmermann, E. H. Brandt, M. Bauer, E. Seider, and L. Genzel, \textit{Physica C} \textbf{183}, 99-104 (1991).
%
\bibitem{Tinkham 1996} M. Tinkham, "Introduction to Superconductivity", 2nd edn. (Dover Publications, 1996).
%
\bibitem{Dressel 2002} M. Dressel, G. Gr\"uner, "Electrodynamics of Solids" (Cambridge University Press, Cambridge, 2002).
%
\bibitem{Tu 2010} J. J. Tu, J. Li, W. Liu, A. Punnoose, Y. Gong, Y. H. Ren, L. J. Li, G. H. Cao, Z. A. Xu, and C. C. Homes, \textit {Phys. Rev. B} \textbf {82}, 174509 (2010).
%
\bibitem{Xing 2018} Zhen Xing, Shanta Saha, J. Paglione, and M. M. Qazilbash, \textit {Phys. Rev. B} \textbf {98}, 224505 (2018).
%
\bibitem{Maksimov} E. G. Maksimov, A. E. Karakozov, B. P. Gorshunov, A. S. Prokhorov, A. A. Voronkov,  E. S. Zhukova, V. S. Nozdrin, S. S. Zhukov, D. Wu, M. Dressel, S. Haindl, K. Iida, and B. Holzapfel, \textit{Phys. Rev. B} \textbf{83}, 140502(R) (2011).
%
\bibitem{Dai 2013} Y. M. Dai, B. Xu, B. Shen, H. Xiao, H. H. Wen, X. G. Qiu, C. C. Homes, and R. P. S. M. Lobo, \textit {Phys. Rev. Lett.} \textbf {111}, 117001 (2013).
%
\bibitem{Lee 2015} Seokbae Lee, Ki-Young Choi, Eilho Jung, Seulki Roh, Soohyeon Shin, Tuson Park, and Jungseek Hwang, \textit {Sci. Rep.} \textbf {5}, 12156 (2015).
%
\bibitem{Abrikosov 1963} A. A. Abrikosov, L. P. Gor'kov, and I. Y. Dsyaloshinskii, "Methods of Quantum Field Theory in Statistical Physics" (Prentice-Hall, Englewood, 1963).
%
\bibitem{Pines 1966} D. Pines and P. Nozi\`eres, "The Theory of Quantum Liquids" (Addison-Wesley, Reading, 1966), Vol. 1.
%
\bibitem{Barisic 2010} N. Bari\u{s}i\'{c}, D. Wu, M. Dressel, L. J. Li, G. H. Cao, and Z. A. Xu, \textit{Phys. Rev. B} \textbf{82}, 054518 (2010).
%
\bibitem{Allen} P. B. Allen, \textit{Phys. Rev. B} \textbf{17}, 3725-34 (1978).
%
\bibitem{Grimvall} G. Grimvall, "The electron-phonon interaction in metals" (North-Holland, 1981).
%
\bibitem{Allen 1981} P. B. Allen and B. Chakraborty, \textit{Phys. Rev. B} \textbf{23}, 4815-27 (1981).
%
\bibitem{Golubov 2011} A. A. Golubov, O. V. Dolgov, A. V. Boris, A. Charnukha, D. L. Sun, C. T. Lin, A. F. Shevchun, A. V. Korobenko, M. R. Trunin, and V. N. Zverev, \textit{JETP Lett.} \textbf{94}, 333--337 (2011); A. Charnukha, O. V. Dolgov, A. A. Golubov, Y. Matiks, D. L. Sun, C. T. Lin, B. Keimer, and A. V. Boris, \textit{Phys. Rev. B} \textbf{84}, 174511 (2011).
%
\bibitem{resumma} G. A. Ummarino, S. Galasso, P. Pecchio, D. Daghero, R. S. Gonnelli, F. Kurth, K. Iida, and B. Holzapfel, \textit{Phys. Status Solidi B} \textbf{252}, No. 4, 821-827 (2015).
%
\bibitem{Lilia} L. Boeri, O. V. Dolgov and A. A. Golubov, \textit{Phys. Rev. Lett.} \textbf{101}, 026403 (2008).
%
\bibitem{Popovich} P. Popovich, A. V. Boris, O. V. Dolgov, A. A. Golubov, D. L. Sun, C. T. Lin, R. K. Kremer, and
B. Keimer, \textit{Phys. Rev. Lett. }\textbf{105}, 027003 (2010).
%
\bibitem{Inosov} D. S. Inosov, J. T. Park, P. Bourges, D. L. Sun, Y. Sidis, A. Schneidewind, K. Hradil, D. Haug, C. T. Lin, B. Keimer, and V. Hinkov, \textit{Nature Physics} \textbf{6}, 178-181 (2010).
%
\bibitem{Carbi} D. Wu, N. Bari\u{s}i\'{c}, M. Dressel, G. H. Cao, Z-A. Xu, E. Schachinger, and J. P. Carbotte, \textit{Phys. Rev. B} \textbf{82}, 144519 (2010).
%
\bibitem{Maxi} E. G. Maksimov, M. L. Kulic, and O. V. Dolgov, \textit{Advances in Condensed Matter Physics}, \textbf{2010}, 1-65 (2010).
}
\end{thebibliography}
\end{document}


\title{Supplementary material for: Optical and Hidden Transport Properties of BaFe$_{1.91}$Ni$_{0.09}$As$_{2}$ Film}

\author{Yu.A. Aleshchenko$^1$}
\email{*E-mail:aleshchenkoya@lebedev.ru}
\author{A.V. Muratov$^1$}
\author{G.A. Ummarino$^{2,3}$}
\author{S. Richter$^{4,5}$}
\author{A. Anna Thomas$^{4,5}$}
\author{R.H\"uhne$^4$}
\affiliation{$^1$ P.N. Lebedev Physical Institute, Russian Academy of Sciences, Leninskiy Prospekt 53, Moscow 119991, Russia}
\affiliation{$^2$ Istituto di Ingegneria e Fisica dei Materiali, Dipartimento di Scienza Applicata e Tecnologia, Politecnico di
Torino, Corso Duca degli Abruzzi 24, 10129 Torino, Italy}
\affiliation{$^3$ National Research Nuclear University MEPhI (Moscow Engineering Physics Institute), Kashirskoe shosse 31, Moscow 15409, Russia}
\affiliation{$^4$ Institute for Metallic Materials, Leibniz IFW Dresden, Helmholtzstrasse 20, Dresden 01069, Germany}
\affiliation{$^5$ School of Sciences, TU Dresden, 01062 Dresden, Germany}

\maketitle
\section{EXPERIMENTAL DETAILS}
The reflectivity of the system film/substrate as well as the optical functions calculated from it depend on the film thickness $d$ [see Eq. (S5) for the two-layer system]. For this reason, the model parameters (Tables SIII--SV) turn out to be dependent on $d$. The calculated normal-state resistivity of the film proves to be also dependent on $d$ since the model plasma frequencies are used in the calculations within the Eliashberg formalism. Thus one should know this parameter when modeling the optical properties of the film. The thickness of the film deposited in the PLD process was checked from time to time using focused ion beam (FIB) cuts. In general, the laser energy during deposition is carefully controlled, which determines the growth rate of the film. This energy is measured inside the chamber for each film. If we use the same energy, the same spot size and the same pulse number, we assume a similar thickness for the resulting film. Nevertheless, there are always slight deviations, which gives the error bar of $\pm 10$~nm.

For IR measurements a BaFe$_{1.91}$Ni$_{0.09}$As$_2$ film with an approximate thickness of $120\pm 10$~nm on the square (001) CaF$_2$ substrate $10\times 10\times 0.46$~mm$^3$ was mounted to the cold finger of the Konti Spectro A continuous-flow cryostat. The motorized sample holder of this cryostat allowed reproducible interchange of the sample and the gold reference mirror under vacuum with minimal alignment variation. A proper alignment of the film with respect to the reference mirror was checked with radiation of a diode laser. Optical conductivity measurements were carried out in a Bruker IFS 125HR Fourier transform infrared (FTIR) spectrometer in near normal reflectance ($\approx~11^\circ $) and normal transmittance geometries with photon energies from 3 meV to 1.25 eV. A systematic error in the reflectance measurements in this region was within 2$\% $. At lower energies, strong oscillations prevent accurate measurements and set the lower limit in our experiment. Large apertures cause artefacts below 3~meV in the experimental spectra, therefore this aperture was set not higher than 4~mm. The experimental configurations for IR measurements including the material of the cryostat windows are presented in Table~SI.

\begin{table}
\caption {The experimental configurations for IR measurements.}
\begin{tabular}{|c|c|c|c|c|c|}
\hline
Range & Source & Beam & Window & Detector & Aperture\\
cm$^{-1}$ & & splitter & & & mm \\
\hline
NIR & Tungsten &  CaF$_2$ & KRS-5 & DLaTGS & 3.15 \\
1200- & lamp & & & & \\
10000 & & & & & \\
\hline
MIR & Globar & KBr & KRS-5 & DLaTGS & 3.15 \\
450-- & & & & & \\
5200 & & & & & \\
\hline
FIR-1 & Hg & Multi- & Mylar & DLaTGS & 3.15 \\
50-- & lamp & layer & & & \\
660 & & Mylar & & & \\
\hline
FIR-2 & Hg & Multi- & Mylar & Liquid & 4 \\
25-- & lamp & layer & & He-cooled & \\
690 & & Mylar & & Si & \\
& & & & bolometer & \\
\hline
\end{tabular}
\end{table}
Additionally, variable-angle spectroscopic ellipsometry was perfomed from 0.5 to 6.2~eV (4000--50000~cm$^{-1}$) with a Woollam VASE ellipsometer in a high-vacuum Janis CRF 725V cryostat (4.2--300 K).

\section{OPTICAL PROPERTIES OF THE CaF$_2$ SUBSTRATE}
The reflectivity and transmittance spectra of the bare CaF$_2$ substrate, which was optically polished from both sides, were measured as a first step [see Fig.~S1(a),(c)]. The transmission bands of CaF$_2$ lie below 250~cm$^{-1}$ and above 600~cm$^{-1}$. The refractive index $n$ and the extinction coefficient $\kappa $ of the substrate in its transmission bands were determined by solving a system
\begin{equation}
\left\{
\begin{matrix}
R(n,\kappa )=R_{exp},\\
T(n,\kappa )=T_{exp}.
\end{matrix}\right.
\end{equation}
Here $R$ ($R_{exp}$) and $T$ ($T_{exp}$) are the model (experimental) reflectance and transmittance, respectively. Given the complex refractive index $\tilde N=n+i\kappa $ of CaF$_2$, we can calculate the reflectance of the bulk material $R_{bulk}$ using the formula
\begin{equation}
R=\left |\frac {1-\tilde N}{1+\tilde N}\right |^2=\frac {(1-n)^2+\kappa ^2}{(1+n)^2+\kappa ^2}.
\end{equation}
The calculated reflectance $R_{bulk}$ was joined together with the experimental reflectance within the range of 250--600~cm$^{-1}$ to obtain $R_{bulk}$ in the whole spectral range [Fig.~S1(b)]. The complex optical conductivity of the CaF$_2$ substrate was calculated from the reflectivity spectrum using the software RefFIT~\cite {Kuzmenko 2018}, which utilizes a multi-oscillator fit of the reflectivity data anchored by the dielectric function $\tilde\varepsilon (\omega ) =\varepsilon _1(\omega )+i\varepsilon _2(\omega )$ measured through ellipsometry~\cite{Kuzmenko 2005,Qazilbash 2009}.
\begin{figure*}[t]
\includegraphics{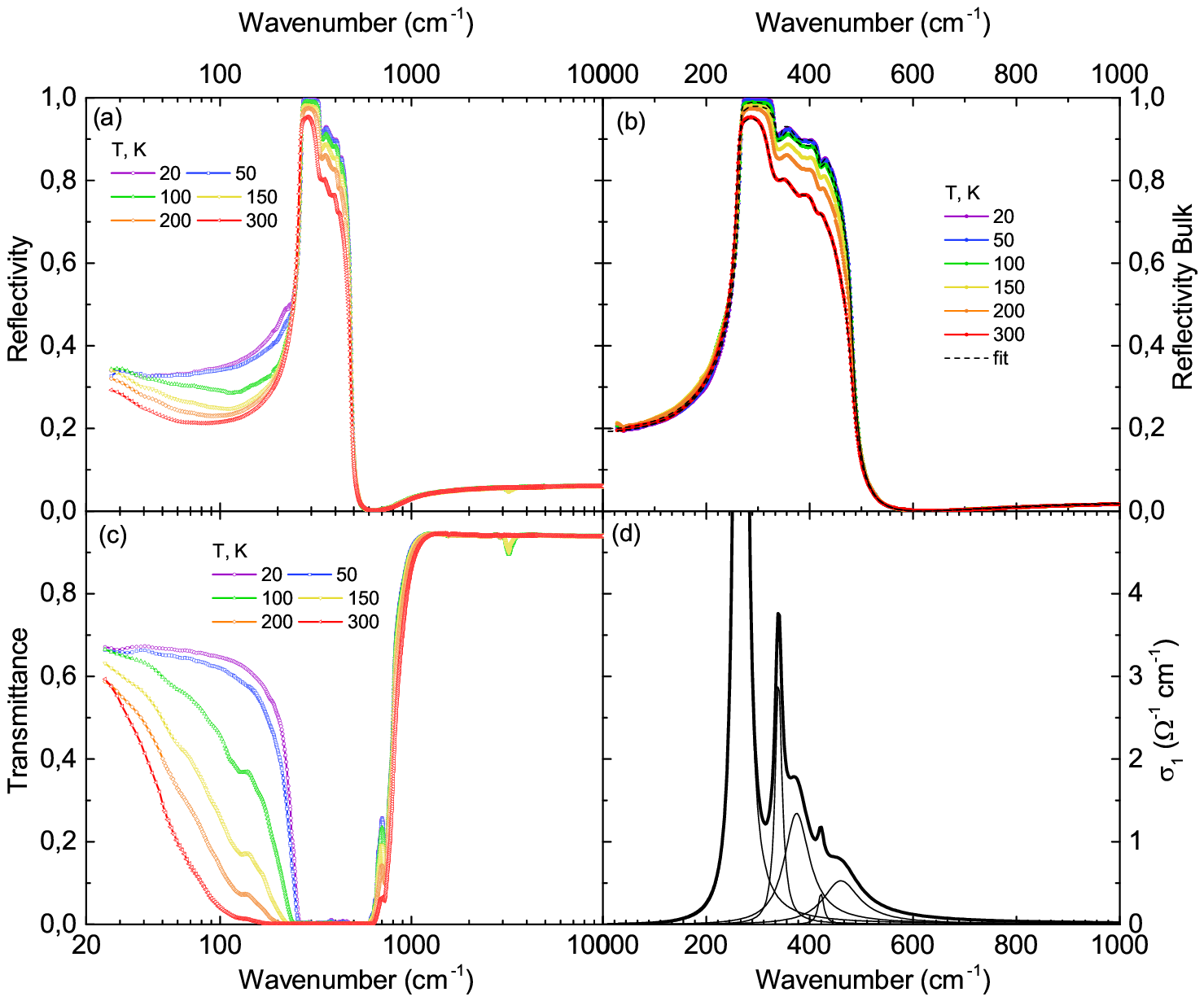}
\caption{(Color online) The measured reflectivity (a) and transmittance (c) spectra of the bare CaF$_2$ substrate at various temperatures. (b) The calculated reflectivity spectrum of the bulk CaF$_2$. (d) Optical conductivity of CaF$_2$ (thick curve) at 20~K together with the constituent Lorentzians (thin lines).}
\end{figure*}

The dielectric functions for each temperature were fitted with five Lorentz oscillators (one of them for modeling the CaF$_2$ phonon contribution at $\sim 260$~cm$^{-1}$):
\begin{equation}
\begin{matrix}
\varepsilon =\varepsilon _{\infty }+\sum\limits_{i=1}^5 \frac {\omega _{i,p}^2}{\omega _{0,i}-\omega ^2-i\omega\gamma _i},\\
\omega _{i,p}=\sqrt{\Delta\varepsilon _i}\omega _{0,i},
\end{matrix}
\end{equation}
where $\varepsilon _\infty $ is the background dielectric function, which comes from contribution of the high frequency absorption, $\Delta\varepsilon _i$ is the oscillator strength. Each $i$th Lorentz oscillator is characterized by its plasma frequency $\omega _{i,p}$, natural frequency $\omega _{0,i}$, and linewidth $\gamma _i$. The values of these fitting parameters are listed in Table~SII. The real part of the optical conductivity $\tilde\sigma = \sigma _1+i\sigma _2$ can then be calculated by $\sigma _1=\omega\varepsilon _2/4\pi $ [Fig. S1(d)].
\begin{table*}
\begin{center}
\caption{The fitting papameters for CaF$_2$.}
\begin{tabular}{|c|c|c|c|c|c|c|c|}
\hline
Contribution & Parameter & 300 K & 200 K & 150 K & 100 K & 50 K & 20 K\\
& & & & & & & (5--30 K)\\
\hline
$E_{inf}$& $\varepsilon _{\infty }$ & 2.05 & 2.11 & 2.13 & 2.11 & 2.06 & 2.06\\
\hline
Lor 1 & $\Delta\varepsilon _1$ & 4.33 & 4.57 & 4.65 & 4.63 & 4.47 & 4.44\\
\hhline{|~|-|-|-|-|-|-|-|}
& $\omega _{0,1}$, cm$^{-1}$ & 260 & 263 & 264 & 265 & 266 & 266\\
\hhline{|~|-|-|-|-|-|-|-|}
& $\gamma _1 $, cm$^{-1}$ & 4.56 & 3.25 & 2.80 & 2.10 & 1.37 & 1.05\\
\hline
Lor 2 & $\Delta\varepsilon _2$ & 0.106 & 0.057  & 0.041 & 0.029 & 0.025 & 0.025 \\
\hhline{|~|-|-|-|-|-|-|-|}
& $\omega _{0,2}$, cm$^{-1}$ & 333 & 338 & 340 & 341 & 340 & 340\\
\hhline{|~|-|-|-|-|-|-|-|}
& $\gamma _2 $, cm$^{-1}$ & 39.1 & 27.0 & 22.6 & 18.9 & 16.8 & 16.6\\
\hline
Lor 3 & $\Delta\varepsilon _3$ & 0.058 & 0.045 & 0.039 & 0.034 & 0.032 & 0.034\\
\hhline{|~|-|-|-|-|-|-|-|}
& $\omega _{0,3}$, cm$^{-1}$ & 372 & 374 & 375 & 375 & 374 & 375\\
\hhline{|~|-|-|-|-|-|-|-|}
& $\gamma _3 $, cm$^{-1}$ & 60.3 & 57.3 & 57.9 & 58.9 & 57.5 & 59.4\\
\hline
Lor 4 & $\Delta\varepsilon _4$ & 0.0014 & 0.0010 & 0.0011 & 0.0012 & 0.0014 & 0.0017\\
\hhline{|~|-|-|-|-|-|-|-|}
& $\omega _{0,4}$, cm$^{-1}$ & 414 & 418 & 420 & 421 & 422 & 422\\
\hhline{|~|-|-|-|-|-|-|-|}
& $\gamma _4 $, cm$^{-1}$ & 14.6 & 10.3 & 10.4 & 11.3 & 11.7 & 13.5\\
\hline
Lor 5 & $\Delta\varepsilon _5$ & 0.035 & 0.028 & 0.021 & 0.017& 0.016 & 0.013\\
\hhline{|~|-|-|-|-|-|-|-|}
& $\omega _{0,5}$, cm$^{-1}$ & 439 & 444 & 451 & 457 & 458 & 460\\
\hhline{|~|-|-|-|-|-|-|-|}
& $\gamma _5 $, cm$^{-1}$ & 131 & 120 & 108 & 104 & 101 & 86\\
\hline
\end{tabular}
\end{center}
\end{table*}

\section{ANALYSIS OF THE TWO-LAYER SYSTEM}
For the analysis of the experimental reflectivity spectra it is necessary to consider a two-layer system (BaFe$_{1.91}$Ni$_{0.09}$As$_2$ film--CaF$_2$ substrate) and to evaluate its optical response~\cite{Dressel 2002} from the $\tilde N$ and thickness of each layer. In this case, the power reflection coefficient is
\begin{equation}
R=\left |\tilde r_{1234}\right |^2,
\end{equation}
where $\tilde r_{1234}$ is the complex reflection coefficient for light traveling from vacuum (medium 1) through film (medium 2) and substrate (medium 3) again to vacuum (medium 4),
\begin{equation}
\begin{split}
\tilde r_{1234}=\frac{\tilde r_{12}+\tilde r_{234}exp(-4\pi\tilde N_fkd_f)}{1+\tilde r_{12}\tilde r_{234}exp(-4\pi\tilde N_fkd_f)},\\
\tilde r_{234}=\frac{\tilde r_{23}+\tilde r_{34}exp(-4\pi\tilde N_skd_s)}{1+\tilde r_{23}\tilde r_{34}exp(-4\pi\tilde N_skd_s)},\\
\tilde r_{ij}=\frac{\tilde N_i-\tilde N_j}{\tilde N_i+\tilde N_j}.
\end{split}
\end{equation}
In Eqs. (4) and (5) the indices 1,2,3,4 are the numbers of the media, $\tilde N_f$ and $\tilde N_s$ are the complex refractive indices of the film and substrate, respectively, $d_f=120$~nm and $d_s=460$~$\mu m$ are the respective thicknesses of the film and substrate, $k$ is the wavenumber.

Frequency dependence of the ellipsometric coefficients $\Psi $ and $\Delta $ for the two-layer system BaFe$_{1.91}$Ni$_{0.09}$As$_2$ film--CaF$_2$ substrate is shown in Fig.~S2(a),(b) for three angles of incidence (65$^\circ $, 70$^\circ $, and 75$^\circ $) and temperatures 5 and 300~K. The curves are nearly temperature independent at least above 10000~cm$^{-1}$, so do the ellipsometric parameters of the bare CaF$_2$ substrate, which is transparent in this spectral range. The pseudodielectric function  $\tilde\varepsilon (\omega)=\varepsilon _1(\omega)+i\varepsilon _2(\omega)$ for each wavenumber was calculated using the WASE software in such a way to describe simultaneously the ellipsometric coefficients taken at three angles of incidence. The results are shown in Fig.~S2(c),(d). Next we adjust the $ab$-plane IR reflectance in the range 2500--6000~cm$^{-1}$ to match the reflectance generated from ellipsometric coefficients.

\section{TWO-DRUDE ANALYSIS}
The reflectivity spectra of the BaFe$_{1.91}$Ni$_{0.09}$As$_2$ film were fitted to the Drude-Lorentz model using the software RefFIT. In the two-Drude approach ~\cite{Wu 2010} the optical conductivity in the normal state is modeled by two Drude components, one narrow ($D_1$) and another broad one ($D_2$), and by a set of Lorentz components representing the response associated with
localized charges and/or optical interband transitions. The complex dielectric function $\tilde\varepsilon (\omega)=\varepsilon _1(\omega)+i\varepsilon _2(\omega)$ can be written as
\begin{equation}
\begin{split}
\tilde\varepsilon (\omega)=\varepsilon _{\infty }-\sum _{i=1}^2\frac{\omega _{Di,p}^2}{\omega (\omega +i\gamma _{Di})}\\
+\sum _{j=1}^5\frac{\omega _{j,p}^2}{\omega _{0,j}^2-\omega ^2-i\gamma _j\omega},
\end{split}
\end{equation}
where $D_1$ ($D_2$) stands for the narrow (broad) Drude component, $\omega _{Di,p}=\sqrt {4\pi\sigma _{0,i}\gamma _{Di}}$ is the Drude plasma frequency, $\gamma _{Di}$ is the (average) elastic scattering rate among free charge carriers, $\omega _{j,p}=\sqrt {\Delta\varepsilon _{0j}}\omega _{0j}$, $\omega _{0,j}$, and $\gamma _j$ are the plasma frequency, the center frequency, and the width of the $j$th Lorentz component, respectively. We use five Lorentz oscillators, one for modeling the phonon at $\sim 94$~cm$^{-1}$ of BaFe$_{1.91}$Ni$_{0.09}$As$_2$, the rest for the contributions of the high-frequency interband transitions. The phonon at $\sim 260$~cm$^{-1}$ observed in Ba122 compounds is much weaker than the strongest phonon peak of the CaF$_2$ substrate at the same frequency. For this reason, it was not included into the film model, which, nevertheless, provided a good quality of the fit. The optical conductivity can be related to the dielectric function as $\tilde\sigma (\omega )=\sigma _1(\omega)+i\sigma _2(\omega )=i\omega [\varepsilon _\infty -\tilde\varepsilon (\omega )]/4\pi $. The fitting parameters for the two-Drude model of the BaFe$_{1.91}$Ni$_{0.09}$As$_2$ film in the normal state are presented in Table SIII. The results of the two-Drude modeling are shown in Fig. 2 of the main paper. In Fig.~3 of the main paper are depicted the temperature dependences of the fitting parameters of the two Drude modes and the resistivities calculated from the optical data.
\begin{table*}
\begin{center}
\caption{The fitting papameters for the two-Drude model of the BaFe$_{1.91}$Ni$_{0.09}$As$_2$ film.}
\begin{tabular}{|c|c|c|c|c|c|c|c|c|c|}
\hline
Contribution & Parameter & 300 K & 200 K & 150 K & 100 K & 50 K & 30 K & 25 K & 20 K\\
\hline
$E_{inf}$& $\varepsilon _{\infty }$ & 1.55 & 1.55 & 1.55 & 1.55 & 1.55 & 1.55 & 1.55 & 1.55\\
\hline
Dr 1 & $\sigma _{0,1}$, $\Omega ^{-1}$cm$^{-1}$ & 3892 & 1801 & 1549 & 1572 & 1499 & 1557 & 1601 & 1621\\
\hhline{|~|-|-|-|-|-|-|-|-|-|}
& $\gamma _{D1}$, cm$^{-1}$ & 845 & 1778 & 1856 & 1787 & 2178 & 2104 & 2107 & 2116\\
\hline
Dr 2 & $\sigma _{0,2}$, $\Omega ^{-1}$cm$^{-1}$ & 1115 & 5609 & 7947 & 9890 & 11658 & 12200 & 12815 & 13113\\
\hhline{|~|-|-|-|-|-|-|-|-|-|}
& $\gamma _{D2}$, cm$^{-1}$ & 256 & 254 & 194 & 148 & 123 & 115 & 107 & 103\\
\hline
Lor 1 & $\Delta\varepsilon _1$ & 6.67 & 6.67 & 6.67 & 6.67 & 6.67 & 6.67 & 6.67 & 6.67\\
\hhline{|~|-|-|-|-|-|-|-|-|-|}
& $\omega _{0,1}$, cm$^{-1}$ & 93.70 & 93.70 & 93.70 & 93.70 & 93.70 & 93.70 & 93.70 & 93.70\\
\hhline{|~|-|-|-|-|-|-|-|-|-|}
& $\gamma _1 $, cm$^{-1}$ & 2.65 & 2.65 & 2.65 & 2.65 & 2.65 & 2.65 & 2.65 & 2.65\\
\hline
Lor 2 & $\Delta\varepsilon _2$ & 67.6 & 37.2 & 43.3 & 44.8 & 47.3 & 41.8 & 37.9 & 36.9\\
\hhline{|~|-|-|-|-|-|-|-|-|-|}
& $\omega _{0,2}$, cm$^{-1}$ & 2011 & 2082 & 1995 & 2059 & 2066 & 2066 & 2029 & 2035\\
\hhline{|~|-|-|-|-|-|-|-|-|-|}
& $\gamma _2 $, cm$^{-1}$ & 3293 & 2668 & 2677 & 2866 & 3192 & 3192 & 2739 & 2722\\
\hline
Lor 3 & $\Delta\varepsilon _3$ & 3.14 & 4.37 & 4.48 & 9.36 & 2.28 & 2.73 & 2.73 & 2.73\\
\hhline{|~|-|-|-|-|-|-|-|-|-|}
& $\omega _{0,3}$, cm$^{-1}$ & 6033 & 6033 & 6106 & 6250 & 6538 & 6211 & 6211 & 6211\\
\hhline{|~|-|-|-|-|-|-|-|-|-|}
& $\gamma _3 $, cm$^{-1}$ & 4749 & 4954 & 4544 & 6117 & 3748 & 3748 & 3748 & 3748\\
\hline
Lor 4 & $\Delta\varepsilon _4$ & 26.9 & 26.9 & 26.9 & 19.5 & 25.0 & 27.1 & 27.1 & 27.1\\
\hhline{|~|-|-|-|-|-|-|-|-|-|}
& $\omega _{0,4}$, cm$^{-1}$ & 15820 & 15820 & 15820 & 17246 & 15879 & 15244 & 15244 & 15244\\
\hhline{|~|-|-|-|-|-|-|-|-|-|}
& $\gamma _4 $, cm$^{-1}$ & 44326 & 44326 & 44326 & 39373 & 41469 & 40639 & 40639 & 40639\\
\hline
Lor 5 & $\Delta\varepsilon _5$ & 1.73 & 1.73 & 1.73 & 1.73 & 1.73 & 1.73 & 1.73 & 1.73\\
\hhline{|~|-|-|-|-|-|-|-|-|-|}
& $\omega _{0,5}$, cm$^{-1}$ & 54207 & 54207 & 54207 & 54207 & 54207 & 54207 & 54207 & 54207\\
\hhline{|~|-|-|-|-|-|-|-|-|-|}
& $\gamma _5 $, cm$^{-1}$ & 36788 & 36788 & 36788 & 36788 & 38209 & 34388 & 34388 & 34388\\
\hline
\end{tabular}
\end{center}
\end{table*}

\section{ONE DRUDE--ONE LORENTZIAN FIT}
The optical conductivity can be decomposed in several ways. The two-Drude analysis presented here, however, is the simplest one. One main argument against a two-Drude analysis is a very large quasiparticle scattering rate of the broad `incoherent' Drude term leading to a unphysically small mean-free path. This difficulty can be completely resolved, if one considers interacting quasiparticles. In this case, the total scattering is a combination of the elastic scattering from impurities in the crystal and the inelastic scattering mediated by various bosonic excitations coupled to the itinerant charge carriers (such as phonons, spin fluctuations, orbital fluctuations, etc.). The inelastic scattering rate is determined by the strength of the electron-boson interaction and is unrelated to the lattice and impurities in the crystal. One of the plausible candidates for the source of the strong interactions underlying the incoherent term in the optical conductivity seems to be Hund's-coupling orbital correlations between Fe-d electrons, predicted to have a sizable intensity and play an important role in the low-energy electrodynamics of the iron-based materials~\cite{Charnukha 2014}.

Other decomposition models for the analysis of the optical conductivity might be used as well. For example, the next simplest one is to fit the low-frequency conductivity by one Drude and one (overdamped) Lorentzian term. The latter replaces the broad Drude component resulting from incoherent background. This fitting procedure was successfully applied previously to describe the low-frequency response of ferropnictides~\cite{Hu 2008,Tu 2010,Xing 2016}. To fit our optical conductivity spectra of the BaFe$_{1.91}$Ni$_{0.09}$As$_2$ film in the normal state, we varied the parameters of a narrow Drude and an overdamped Lorentz components leaving the Lorentzian contributions associated with localized charges and/or optical interband transitions the same as in the two-Drude analysis. The fitting parameters of the one Drude--one Lorentzian fit are tabulated in Table SIV.

\begin{figure*}
\includegraphics{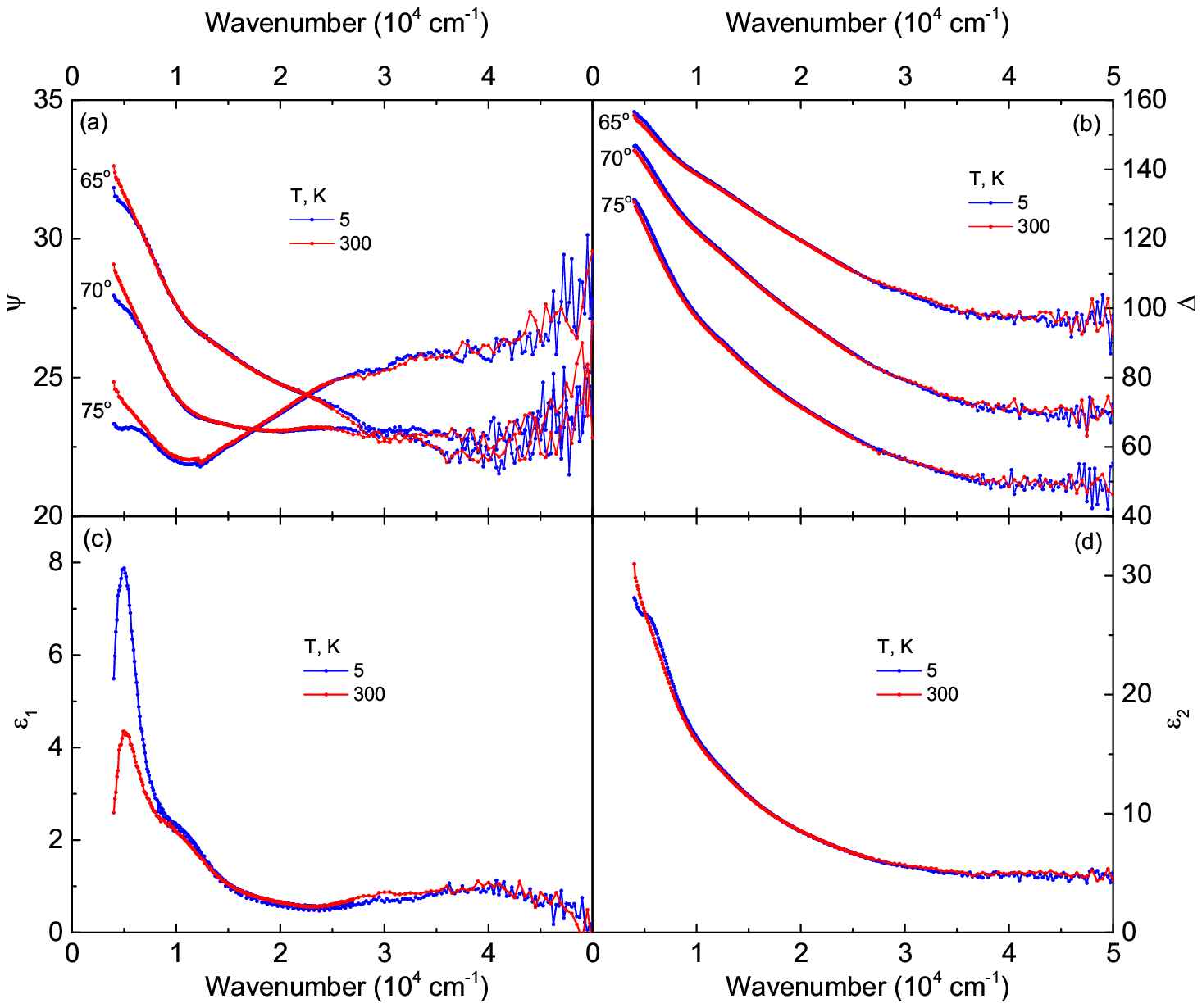}[t]
\caption{(Color online) (a),(b) Frequency-dependent ellipsometric coefficients $\Psi $ and $\Delta $ for the two-layer system BaFe$_{1.91}$Ni$_{0.09}$As$_2$ film--CaF$_2$ for three angles of incidence and temperatures 5 and 300~K; (c),(d) real ($\varepsilon _1$) and imaginary ($\varepsilon _2$) parts of the pseudodielectric function $\tilde\varepsilon $ calculated using the WASE software.}
\end{figure*}

Figure~S3 shows the temperature dependences of the Drude parameters taken from our fits. The plasma frequency [Fig.~S3(a)] is almost temperature independent similar to that determined from the two-Drude analysis [see Fig.~3(a) of the main paper]. The dashed lines in Fig.~S3(b),(d) denote the $T^2$ fits to the scattering rate for the Drude component and to the DC resistivity that implies the charge carriers are described by the Fermi-liquid theory. This conclusion is supported by the results of the two-Drude analysis [Fig.~3(b),(d) of the main paper].
\begin{table*}
\begin{center}
\caption{The fitting papameters for the one Drude--one Lorentzian model of the BaFe$_{1.91}$Ni$_{0.09}$As$_2$ film.}
\begin{tabular}{|c|c|c|c|c|c|c|c|c|c|}
\hline
Contribution & Parameter & 300 K & 200 K & 150 K & 100 K & 50 K & 30 K & 25 K & 20 K\\
\hline
$E_{inf}$& $\varepsilon _{\infty }$ & 1.55 & 1.55 & 1.55 & 1.55 & 1.55 & 1.55 & 1.55 & 1.55\\
\hline
Dr 0 & $\sigma _{0}$, $\Omega ^{-1}$cm$^{-1}$ & 4990 & 7416 & 9396 & 11233 & 13058 & 13348 &  13998 & 14278\\
\hhline{|~|-|-|-|-|-|-|-|-|-|}
& $\gamma _{D0}$, cm$^{-1}$ & 445 & 288 & 218 & 170 & 138 & 133 & 123 & 120\\
\hline
Lor 0 & $\Delta\varepsilon _1$ & 158 & 191 & 201 & 229 & 289 & 284 & 325 & 335\\
\hhline{|~|-|-|-|-|-|-|-|-|-|}
& $\omega _{0,L0}$, cm$^{-1}$ & 707 & 878 & 827 & 781 & 767 & 768 & 733 & 729\\
\hhline{|~|-|-|-|-|-|-|-|-|-|}
& $\gamma _{L0}$, cm$^{-1}$ & 1289 & 2093 & 2052 & 1958 & 2311 & 2182 & 2185 & 2186\\
\hline
Lor 1 & $\Delta\varepsilon _1$ & 6.67 & 6.67 & 6.67 & 6.67 & 6.67 & 6.67 & 6.67 & 6.67\\
\hhline{|~|-|-|-|-|-|-|-|-|-|}
& $\omega _{0,1}$, cm$^{-1}$ & 93.70 & 93.70 & 93.70 & 93.70 & 93.70 & 93.70 & 93.70 & 93.70\\
\hhline{|~|-|-|-|-|-|-|-|-|-|}
& $\gamma _1 $, cm$^{-1}$ & 2.65 & 2.65 & 2.65 & 2.65 & 2.65 & 2.65 & 2.65 & 2.65\\
\hline
Lor 2 & $\Delta\varepsilon _2$ & 67.6 & 37.2 & 43.3 & 44.8 & 47.3 & 41.8 & 37.9 & 36.9\\
\hhline{|~|-|-|-|-|-|-|-|-|-|}
& $\omega _{0,2}$, cm$^{-1}$ & 2011 & 2082 & 1995 & 2059 & 2066 & 2066 & 2029 & 2035\\
\hhline{|~|-|-|-|-|-|-|-|-|-|}
& $\gamma _2 $, cm$^{-1}$ & 3293 & 2668 & 2677 & 2866 & 3192 & 3192 & 2739 & 2722\\
\hline
Lor 3 & $\Delta\varepsilon _3$ & 3.14 & 4.37 & 4.48 & 9.36 & 2.28 & 2.73 & 2.73 & 2.73\\
\hhline{|~|-|-|-|-|-|-|-|-|-|}
& $\omega _{0,3}$, cm$^{-1}$ & 6033 & 6033 & 6106 & 6250 & 6538 & 6211 & 6211 & 6211\\
\hhline{|~|-|-|-|-|-|-|-|-|-|}
& $\gamma _3 $, cm$^{-1}$ & 4749 & 4954 & 4544 & 6117 & 3748 & 3748 & 3748 & 3748\\
\hline
Lor 4 & $\Delta\varepsilon _4$ & 26.9 & 26.9 & 26.9 & 19.5 & 25.0 & 27.1 & 27.1 & 27.1\\
\hhline{|~|-|-|-|-|-|-|-|-|-|}
& $\omega _{0,4}$, cm$^{-1}$ & 15820 & 15820 & 15820 & 17246 & 15879 & 15244 & 15244 & 15244\\
\hhline{|~|-|-|-|-|-|-|-|-|-|}
& $\gamma _4 $, cm$^{-1}$ & 44326 & 44326 & 44326 & 39373 & 41469 & 40639 & 40639 & 40639\\
\hline
Lor 5 & $\Delta\varepsilon _5$ & 1.73 & 1.73 & 1.73 & 1.73 & 1.73 & 1.73 & 1.73 & 1.73\\
\hhline{|~|-|-|-|-|-|-|-|-|-|}
& $\omega _{0,5}$, cm$^{-1}$ & 54207 & 54207 & 54207 & 54207 & 54207 & 54207 & 54207 & 54207\\
\hhline{|~|-|-|-|-|-|-|-|-|-|}
& $\gamma _5 $, cm$^{-1}$ & 36788 & 36788 & 36788 & 36788 & 38209 & 34388 & 34388 & 34388\\
\hline
\end{tabular}
\end{center}
\end{table*}
\begin{figure}
\includegraphics{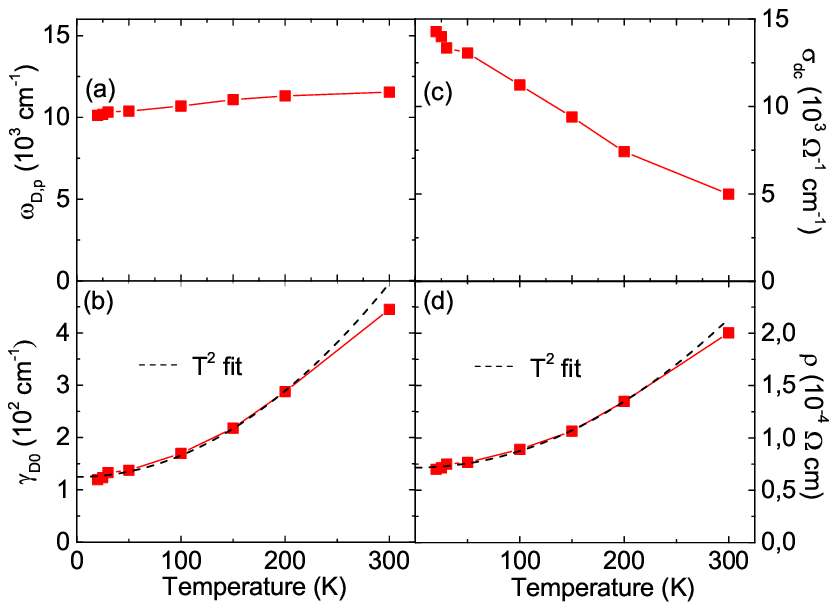}
\caption{(Color online) (a),(b) Frequency-dependent ellipsometric coefficients $\Psi $ and $\Delta $ for the two-layer system BaFe$_{1.91}$Ni$_{0.09}$As$_2$ film--CaF$_2$ for three angles of incidence and temperatures 5 and 300~K; (c),(d) real ($\varepsilon _1$) and imaginary ($\varepsilon _2$) parts of the pseudodielectric function $\tilde\varepsilon $ calculated using the WASE software.}
\end{figure}

\section{SUPERCONDUCTING STATE AND ONE DRUDE--ONE LORENTZIAN MODEL}
To describe the optical conductivity of the BaFe$_{1.91}$Ni$_{0.09}$As$_{2}$ film in the superconducting (SC) state, one should replace the Drude term in the one Drude--one Lorentzian model for the Zimmermann term~\cite{Zimmermann 1991}. For simplicity, the Lorentzian terms are assumed to be the same as those in the normal state. In fitting the conductivity, only a single-band case was initially considered. However, this failed to reproduce the experimental reflectivity spectrum [see the dash-dot line in the inset of Fig.~S4(a)]. In the second step we tried to fit the optical conductivity with two SC gaps. 1) First of all, we divided the Drude conductivity by two equal parts (7139~cm$^{-1}$) and fixed the damping parameter. This approach produces two SC gaps $2\Delta _0^{(1)}=59.2$~cm$^{-1}$ and $2\Delta _0^{(2)}=23.6$~cm$^{-1}$ [see Table~SV]. 2) Then the Drude conductivity was again divided by two equal parts and we varied both the magnitudes of the SC gaps and the damping. In this case we obtain $2\Delta _0^{(1)}=51.5$~cm$^{-1}$ and $2\Delta _0^{(2)}=23.8$~cm$^{-1}$. The results are presented in Table SV and in Fig.~S4(b). These two approaches prove to be virtually identical. It can be shown that the sum of two Drude contributions with the parameters $\sigma _{0,1}=5230$ $\Omega ^{-1}$cm$^{-1}$, $\gamma _{D0,1}=156.5$~cm$^{-1}$ and $\sigma _{0,2}=9410$ $\Omega ^{-1}$~cm$^{-1}$, $\gamma _{D0,2}=94.6$~cm$^{-1}$ is well described by a single contour with $\sigma _{0}=14534$ $\Omega ^{-1}$cm$^{-1}$ and $\gamma _{D0}=114$~cm$^{-1}$. This means that either both gaps are opened in a single band or they are opened in different bands of BaFe$_{1.91}$Ni$_{0.09}$As$_{2}$ having similar parameters. The quality of the fit is demonstrated in Fig.~S4(a). It should be noted that both two-Drude and one Drude--one Lorentzian models give almost similar values for the SC gaps, which increases the reliability of the obtained results.
\begin{table*}
\begin{center}
\caption{The fitting papameters for the one Drude--one Lorentzian model of the BaFe$_{1.91}$Ni$_{0.09}$As$_2$ film in the SC state.}
\begin{tabular}{|c|c|c|c|c|c|c|c|}
\hline
& $\sigma _{0,1}$, & $\gamma _{D0,1}$, & $\sigma _{0,2}$, & $\gamma _{D0,2}$,& $\sigma _{0,1}+\sigma _{0,2} $, & $2\Delta _0^{(1)}$, & $2\Delta _0^{(2)}$,\\
& $\Omega ^{-1}$cm$^{-1}$ & cm$^{-1}$ & $\Omega ^{-1}$cm$^{-1}$ & cm$^{-1}$ &  $\Omega ^{-1}$cm$^{-1}$ & ~cm$^{-1}$ & cm$^{-1}$\\
\hline
1 & 5061 & 120 & 9137 & 120 & 14198 & 59.2 & 23.6\\
\hline
2 & 5230 & 156.5 & 9410 & 94.6 & 14640 & 51.5 & 23.8\\
\hline
\end{tabular}
\end{center}
\end{table*}

\begin{figure}
\includegraphics{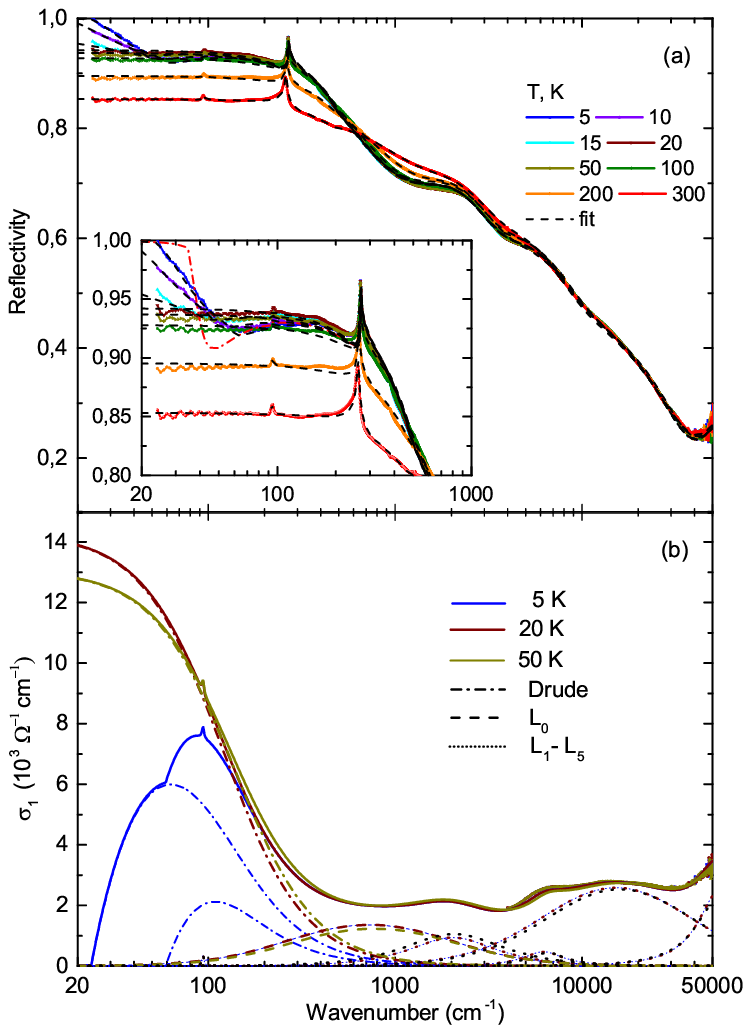}
\caption{(Color online) (a) The measured reflectivity spectra of the BaFe$_{1.91}$Ni$_{0.09}$As$_{2}$ film at various temperatures together with the one Drude--one Lorentzian fit (dashed lines). In the inset we display magnified views of reflectivity spectra for the low frequency region. (b) Three representative Drude-Lorentz fits and optical conductivity data for the BaFe$_{1.91}$Ni$_{0.09}$As$_{2}$ film at temperatures of 5, 20 and 50 K, respectively.}
\end{figure}